\documentclass[aps,prl,twocolumn,showpacs,superscriptaddress,amsmath,amssymb]{revtex4-1}

\usepackage{graphicx}
\usepackage[colorlinks=true,linkcolor=blue,citecolor=blue,urlcolor=blue]{hyperref}
\usepackage{bbold}
\usepackage{gensymb}

\def \d {\delta}
\def \D {\Delta}

\def \ve {\varepsilon}

\def \G{\Gamma}

\def \L {\Lambda}
\def \o {\omega}
\def \O {\Omega}

\def \p {\partial}

\def \del {\nabla}

\newcommand{\intv}[1]{\int_{\mbf #1}}

\newcommand{\sumv}[1]{\sum_{\mbf #1}}

\def \rar {\rightarrow}

\def \la {\langle}
\def \ra {\rangle}
\def \fr {\frac}
\def \lf {\left}
\def \ri {\right}
\newcommand{\bra}[1]{\la#1|}
\newcommand{\ket}[1]{|#1\ra}
\newcommand{\braket}[3]{\la#1|#2|#3\ra}
\newcommand{\innp}[2]{\la#1|#2\ra}

\newcommand{\Braket}[3]{\lf\la#1\lf|#2\ri|#3\ri\ra}

\def \Tr {\mathrm{Tr}}
\def \bece {\begin{center}}
\def \ence {\end{center}}
\def \beeq {\begin{equation}}
\def \eneq {\end{equation}}
\def \beal {\begin{aligned}}
\def \enal {\end{aligned}}
\def \bega {\begin{gathered}}
\def \enga {\end{gathered}}
\def \benu {\begin{enumerate}}
\def \ennu {\end{enumerate}}
\def \beit {\begin{itemize}}
\def \enit {\end{itemize}}
\def \bede {\begin{description}}
\def \ende {\end{description}}
\def \betb {\begin{tabular}}
\def \entb {\end{tabular}}
\def \bear {\begin{array}}
\def \enar {\end{array}}

\def \mbf {\mathbf}

\def \mca {\mathcal}

\def \bsb{\boldsymbol}
\def \txt {\text}

\newcommand{\comment}[1]{}

\begin{document}


\title{Dual Haldane sphere and quantized band geometry in chiral multifold fermions}

\author{Yu-Ping Lin}
\email{Yuping.Lin@colorado.edu}
\affiliation{Department of Physics, University of Colorado, Boulder, Colorado 80309, USA}
\author{Wei-Han Hsiao}
\email{weihanhsiao@uchicago.edu}
\affiliation{Kadanoff Center for Theoretical Physics, University of Chicago, Chicago, Illinois 60637, USA}

\date{\today}

\begin{abstract}
We show that the chiral multifold fermions present a dual Haldane sphere problem in momentum space.  Owing to the Berry monopole at the degenerate point, a dual Landau level emerges in the trace of quantum metric, with which a quantized geometric invariant is defined through a surface integration. We further demonstrate potential manifestations in the measurable, physical observables. With a lower bound derived for the finite spread of Wannier functions, anomalous phase coherence is identified accordingly for the flat band superconductivity. We briefly comment on the stability of these results under perturbations. Potential experimental probes of the quantum metric are also discussed.
\end{abstract}

\maketitle

\begin{figure}[b]
\centering
\includegraphics[scale = 1]{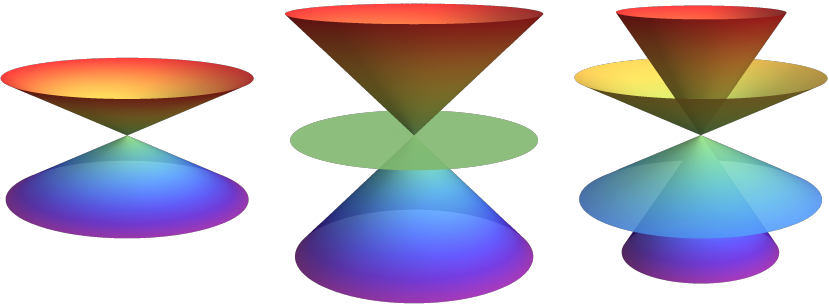}
\caption{\label{fig:cmp} Energy dispersions of chiral multifold fermions with spins $s=1/2$ (left), $1$ (center), and $3/2$ (right).}
\end{figure}

Chiral multifold fermions have received much attention of modern condensed matter research in recent years \cite{armitage18rmp,bradlyn16sc,isobe16prb,tang17prl,boettcher20prl,rao19n,sanchez19n,schroter19np,takane19prl,lv19prb,Schroter20sc}. These fermions possess a topologically protected degenerate point in momentum space (Fig.~\ref{fig:cmp}), with the low-energy theory exhibiting an effective spin-momentum locking. Such structure generically hosts nontrivial topological features in the eigenstates, which originate from the Berry monopole at the degenerate point. The chiral multifold fermions potentially own a variety of novel characteristics. These include the exotic superconductivity with unconventional pairing and/or dramatic enhancement from flat bands \cite{lin18prb,sim19ax,lin20prr,link20prb}, as well as the novel optical responses \cite{flicker18prb,sanchezmartinez19prb}. The study of materials with prototypical spin-$1/2$ chiral multifold fermions, known as the Weyl semimetals, has developed as one of the mainstreams in modern condensed matter research in the past decade \cite{armitage18rmp}. Meanwhile, some realizations with higher spins, including spin-$1$ and $3/2$ fermions, have also been proposed and uncovered in the solid state materials \cite{bradlyn16sc,isobe16prb,tang17prl,rao19n,sanchez19n,schroter19np,takane19prl,lv19prb,Schroter20sc}. Moreover, the realm of higher-spin systems can potentially also be explored with ultracold atomic systems through the coupling of atomic hyperfine states \cite{hu18prl,zhu17pra}.

Compared to the extensively studied topological properties of eigenstates, the `geometric' aspects have not been explored in depth. For an eigenstate dependent on a set of adiabatic parameters, the variation in parameter space manifests in both the phase and the state \cite{berry89}. The phase variation is known as the Berry phase \cite{berry84rspa}, which contributes to the topological properties of the eigenstate \cite{xiao2010rmp}. Meanwhile, the state variation reflects the `quantum distance' between the initial and final states under the parameter change, and is captured by the quantum metric \cite{provost80cmp,page87pra,anandan90prl}. Such a feature can be experimentally probed \cite{bleu18prb,ozawa18prb,asteria19np,klees20prl,yu19nsr,tan19prl,gianfrate20n} by, for example, a periodic drive measurement. The quantum metric manifests on a variety of occasions, including the finite spread of Wannier functions \cite{marzari97prb,matsuura10prb,marzari12rmp}, the anomalous superfluid stiffness on flat bands \cite{peotta15nc,liang17prb,hu19prl,xie20prl}, the realization of fractional Chern insulators \cite{roy14prb,claassen15prl}, the geometric contribution to orbital susceptibility \cite{gao15prb,piechon16prb}, the current noise \cite{neupert13prb}, the indication of phase transition \cite{zanardi07prl,ma10prb,kolodrubetz13prb}, and the measurement of tensor monopoles \cite{palumbo18prl}. Despite the broad range of proposed applications, the inherent properties of quantum metric have not been investigated elaborately.

In this Letter, we aim to understand the inherent properties of quantum metric for the chiral multifold fermions. We show that a dual Haldane sphere \cite{haldane83prl} problem emerges in the computation of the trace of quantum metric owing to the Berry monopole in momentum space. A quantized geometric invariant can be defined through a surface integration, and together with the Chern number it establishes a sum rule. We further demonstrate the potential manifestations of such quantized band geometry in the measurable physical observables. A lower bound is derived for the finite spread of Wannier functions, which can trigger anomalous phase coherence in the flat band superconductivity. We briefly comment on the stability of these results under perturbations. Potential probes of quantum metric in the experimental systems are also discussed.

\begin{figure}[t]
\centering
\includegraphics[scale = 0.2]{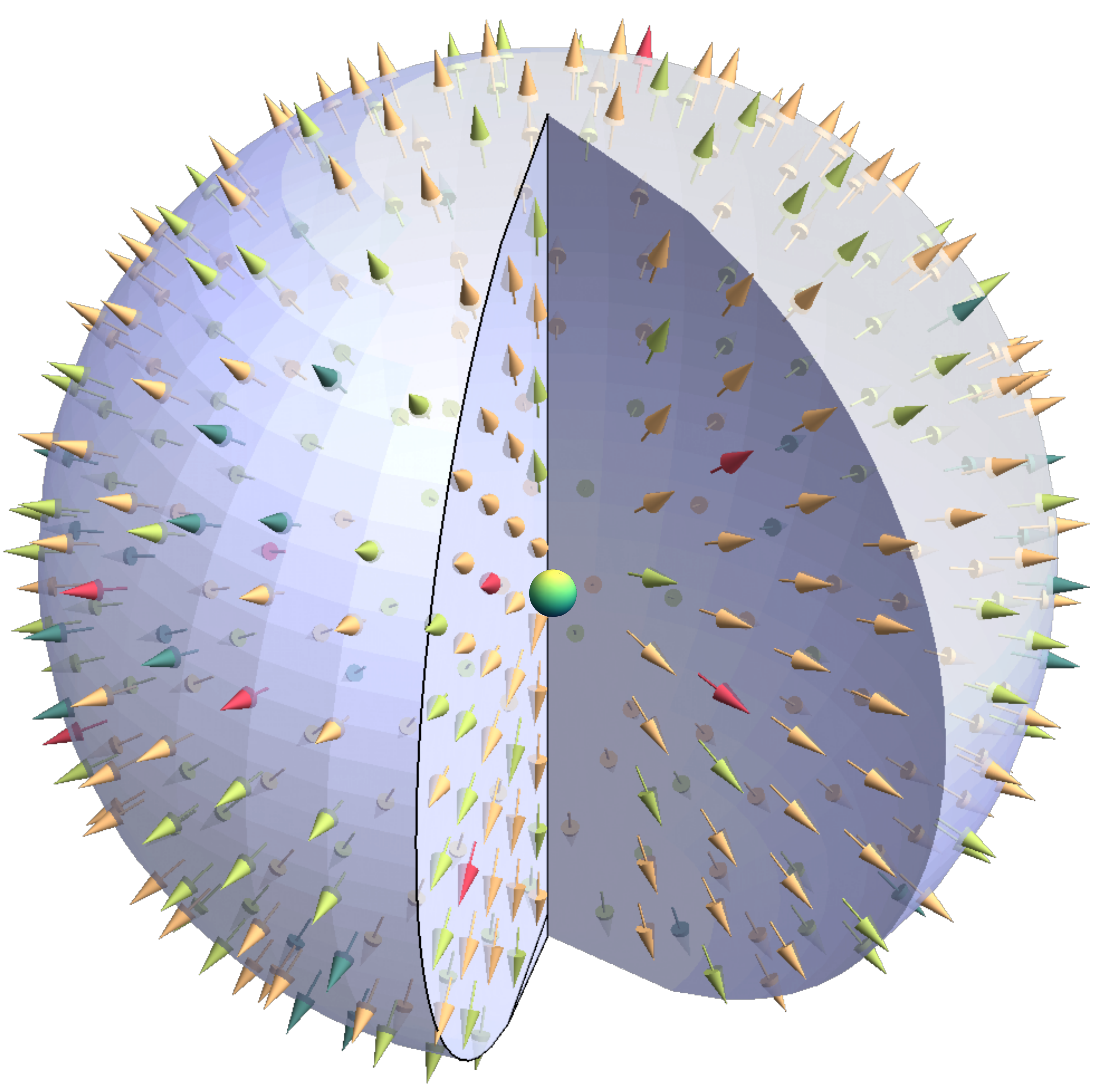}
\caption{\label{fig:magmnp} A monopole emits radial fluxes which are perpendicular to the spherical shells around it. Such configuration illustrates the Berry monopole charge and Berry fluxes for the chiral multifold fermions, as well as the Dirac magnetic monopole and magnetic fields in the Haldane sphere.}
\end{figure}

We begin by introducing a minimal model of chiral multifold fermions (CMF) in three-dimensional (3D) systems. For spin-$s$ fermion with integer or half-integer spin $s=1/2,1,3/2,\dots$, the Hamiltonian reads
\beeq
\label{eq:ham0}
\mca H_{\txt{CMF},\mbf k}=v\mbf k\cdot\mbf S.
\eneq
Here $v$ is the effective velocity, $\mbf k=k\mbf{\hat k}$ denotes the momentum with magnitude $k$ along the direction of unit vector $\mbf{\hat k}$, and $\mbf S$ represents the vector of spin-$s$ operators $S^a$'s with $a=x,y,z$. The Hamiltonian exhibits $2s+1$ eigenstates $\ket{u^{sn}_{\mbf k}}$ with energies $\ve^{sn}_{\mbf k}=vkn$, $n=-s,-s+1,\dots,s$ (Fig.~\ref{fig:cmp}). These eigenstates are nondegenerate except at the degenerate point $\mbf k=\mbf0$. Nontrivial topological properties are generically manifest in these eigenstates. For the $n$-th eigenstate $\ket{u^{sn}_{\mbf k}}$, the calculation of Berry flux $\mbf B^{sn}_{\mbf k}=i\sum_{m\neq n}\braket{u^{sm}_{\mbf k}}{\bsb\del_{\mbf k}\mca H_{\txt{CMF},\mbf k}}{u^{sn}_{\mbf k}}^*\times\braket{u^{sm}_{\mbf k}}{\bsb\del_{\mbf k}\mca H_{\txt{CMF},\mbf k}}{u^{sn}_{\mbf k}}/(\ve^{sm}_{\mbf k}-\ve^{sn}_{\mbf k})^2$ yields $\mbf B^{sn}_{\mbf k}=-(n/k^2)\mbf{\hat k}$ \cite{berry84rspa}.  An integration over any arbitrary closed surface around the degenerate point leads to an integer Chern number $C^{sn}=(1/2\pi)\oint d\mbf S_{\mbf k}\cdot\mbf B^{sn}_{\mbf k}=-2n$ \cite{xiao2010rmp}. The integer Chern number is a topological invariant of the eigenstate $\ket{u^{sn}_{\mbf k}}$. Such an integer corresponds to a quantized monopole charge $q^{sn}=C^{sn}/2=-n$ of Berry flux $\mbf B^{sn}_{\mbf k}$ (Fig.~\ref{fig:magmnp}), which is a momentum-space analogy of Dirac's quantized magnetic monopole \cite{dirac31rspa}. The wavefunction of the eigenstate $\ket{u^{sn}_{\mbf k}}$ can be identified based on symmetry. Since the Hamiltonian (\ref{eq:ham0}) manifests a spin-orbit-coupled rotation symmetry, the wavefunction follows the same symmetry and takes the form $\ket{u^{sn}_{\mbf k}}=\sqrt{4\pi}\sum_{m=-s}^s\innp{ss;-mm}{00}Y_{-q^{sn}s-m}(\mbf{\hat k})\ket{sm}$. Here $\innp{ls;m_lm_s}{jm_j}$ is the Clebsch-Gordan coefficient with orbital, spin, total angular momenta $l,s,j$ and according axial components $m_{l,s,j}$. $Y_{qlm}(\mbf{\hat k})$ denotes the monopole harmonics in the monopole-$q$ sector \cite{wu76npb,wu77prd}, and $\ket{sm}$ stands for the $S^z$ eigenstate. Given $\innp{ss;-mm}{00}=(-1)^{m+s}/\sqrt{2s+1}$, the wavefunction can be expressed in a more convenient form $\ket{u^{sn}_{\mbf k}}=\sqrt{4\pi/(2s+1)}\sum_{m=-s}^sY_{q^{sn}sm}^*(\mbf{\hat k})\ket{sm}$. With the wavefunction in hand, the Berry connection $\mbf A^{sn}_{\mbf k}=\braket{u^{sn}_{\mbf k}}{i\bsb\del_{\mbf k}}{u^{sn}_{\mbf k}}$ can be derived, yielding consistent Berry flux $\mbf B^{sn}_{\mbf k}=\bsb\del_{\mbf k}\times\mbf A^{sn}_{\mbf k}$ and Chern number $C^{sn}=2q^{sn}=-2n$ with previous discussions.

For the eigenstate $\ket{u^{sn}_{\mbf k}}$ of the chiral multifold fermion model (\ref{eq:ham0}), the variation under momentum change is measured by the quantum geometric tensor $T^{sn}_{ab\mbf k}=\braket{\p_{k_a}u^{sn}_{\mbf k}}{(1-\ket{u^{sn}_{\mbf k}}\bra{u^{sn}_{\mbf k}})}{\p_{k_b}u^{sn}_{\mbf k}}$ \cite{berry89}. While the imaginary part of the tensor corresponds to the Berry flux $B^{sn}_{a\mbf k}=-\ve_{abc}\txt{Im}[T^{sn}_{bc\mbf k}]$, our interest lies in the real part $g^{sn}_{ab\mbf k}=\txt{Re}[T^{sn}_{ab\mbf k}]$, known as the quantum or Fubini-Study metric \cite{provost80cmp,page87pra,anandan90prl}. This quantum metric takes the form
\beeq\beal
\label{eq:qm}
g^{sn}_{ab\mbf k}
&=\fr{1}{2}(\innp{\p_{k_a}u^{sn}_{\mbf k}}{\p_{k_b}u^{sn}_{\mbf k}}+\innp{\p_{k_b}u^{sn}_{\mbf k}}{\p_{k_a}u^{sn}_{\mbf k}})\\
&\quad+\innp{u^{sn}_{\mbf k}}{\p_{k_a}u^{sn}_{\mbf k}}\innp{u^{sn}_{\mbf k}}{\p_{k_b}u^{sn}_{\mbf k}}
\enal\eneq
and measures the `quantum distance' $1-|\innp{u^{sn}_{\mbf k}}{u^{sn}_{\mbf k+d\mbf k}}|^2=g^{sn}_{ab\mbf k}dk_adk_b$ in the Hilbert space. Remarkably, we discover a `quantized' trace
\beeq
\label{eq:trqm}
\Tr g^{sn}_{\mbf k}=\fr{1}{k^2}[s(s+1)-(q^{sn})^2]
\eneq
for the quantum metric of chiral multifold fermions. This is the main result of this Letter. The quantization is determined by both the angular momentum $s$ and the monopole charge $q^{sn}$, and is indifferent to the orientation because of the rotation symmetry. Furthermore, the $k^{-2}$ dependence implies a quantized invariant
\beeq
\label{eq:geominv}
G^{sn}=\fr{1}{2\pi}\oint d\mbf S_{\mbf k}\cdot\mbf{\hat k}\Tr g^{sn}_{\mbf k}=2[s(s+1)-(q^{sn})^2],
\eneq
provided the integral domain encloses the degenerate point. The quantization of $G^{sn}$ originates from the monopole harmonics wavefunction $\ket{u^{sn}_{\mbf k}}$, which is protected by the spin-orbit-coupled rotation symmetry around the degenerate point. We thus uncover a `symmetry-protected geometric invariant' $G^{sn}$ in the chiral multifold fermion model (\ref{eq:ham0}). This geometric invariant is different from the Chern number $C^{sn}$, which is a well-recognized topological invariant. Despite the difference, a `sum rule' of these two invariants can be established from the quantization rule
\beeq
\label{eq:sumrule}
G^{sn}+\fr{(C^{sn})^2}{2}=2s(s+1),
\eneq
where the angular momentum $s$ sets a measure of the sum. Note that the spin-$s$ chiral multifold fermions exhibit the maximal Chern number $|C^{s\pm s}|=2s$. This sets a lower bound for the geometric invariant $G^{sn}\geq|C^{sn}|$, as has been identified for general cases from the positive definiteness of quantum geometric tensor \cite{peotta15nc}.

To understand how the chiral multifold fermions acquire the quantized trace of quantum metric (\ref{eq:trqm}) and the geometric invariant $G^{sn}$ (\ref{eq:geominv}), we consider an alternative expression of the quantum metric (\ref{eq:qm})
\beeq
g^{sn}_{ab\mbf k}=\fr{1}{2}\braket{u^{sn}_{\mbf k}}{\{r^{sn}_a,r^{sn}_b\}}{u^{sn}_{\mbf k}}.
\eneq
Here the position $\mbf r^{sn}=i\bsb\del_{\mbf k}-\mbf A^{sn}_{\mbf k}$ corresponds to the covariant derivative in momentum space, where the Berry connection is involved \cite{claassen15prl}. Notably, the trace of quantum metric $\Tr g^{sn}_{\mbf k}=\braket{u^{sn}_{\mbf k}}{|\mbf r^{sn}|^2}{u^{sn}_{\mbf k}}$ manifests the expectation value of momentum-space Laplacian $|\mbf r^{sn}|^2$. This suggests a `duality' between the trace of quantum metric and the kinetic energy, where the roles of position and momentum are exchanged. Since the eigenstates are composed of monopole harmonics $Y_{q^{sn}sm}^*(\mbf{\hat k})$, the radial part of Laplacian $(\mbf{\hat k}\cdot\mbf r^{sn})^2$ does not contribute, leaving only the angular part $|\mbf r^{sn}|^2_\perp=|\mbf r^{sn}|^2-(\mbf{\hat k}\cdot\mbf r^{sn})^2$ in the trace of quantum metric. The angular part of Laplacian can be related to the `dynamical angular momentum' $\mbf \L^{sn}=\mbf r^{sn}\times\mbf k$ through $|\mbf \L^{sn}|^2=k^2|\mbf r^{sn}|^2_\perp$. This leads to an alternative form of the trace of quantum metric
\beeq
\label{eq:trqmangmomt}
\Tr g^{sn}_{\mbf k}=\Braket{u^{sn}_{\mbf k}}{\fr{|\mbf \L^{sn}|^2}{k^2}}{u^{sn}_{\mbf k}},
\eneq
which captures the `dual energy' from the dynamical angular momentum in the eigenstate $\ket{u^{sn}_{\mbf k}}$.

Amazingly, the trace of quantum metric in chiral multifold fermions is algebraically equivalent to the energy of an electron moving on a sphere enclosing a Dirac magnetic monopole. Such equivalence can be observed from the common structure in the two problems, where a rotationally symmetric system around a monopole is defined. When a Dirac magnetic monopole is present at the center of a sphere (Fig.~\ref{fig:magmnp}), the two-dimensional (2D) electron gas experiences a uniform perpendicular magnetic field, thereby manifests the quantum Hall effect. This implies a Landau level quantization for the eigenstates in the energy spectrum \cite{haldane83prl}. The Hamiltonian describing this `Haldane sphere problem' is $H=|\mbf\L|^2/2m_eR^2$, where $\mbf\L=\mbf R\times(-i\bsb\del+e\mbf A)$ is the dynamical angular momentum, $\mbf R=R\mbf{\hat R}$ is the position at constant radius $R$ along the direction of unit vector $\mbf{\hat R}$, $m_e$ and $-e$ are the electronic mass and charge, and $\mbf A$ is the electromagnetic gauge field. Rotation symmetry enforces the angular momentum $l$ and its axial component $m$ as the good quantum numbers. These quantities then determine the quantized Landau level energy $E_{qlm}=[l(l+1)-q^2]/2m_eR^2$ in the presence of monopole charge $q$ \cite{jainbook,hsiao20prb}. The result immediately suggests an analogous quantization for the trace of quantum metric (\ref{eq:trqmangmomt}) in the chiral multifold fermion model (\ref{eq:ham0}). As `dual Haldane spheres' in momentum space, the chiral multifold fermions manifest the `dual Landau level quantization' (\ref{eq:trqm}), consistent with our previous observation from direct calculation. Note that the eigenstates in the Haldane sphere exhibit the monopole harmonics wavefunction $\psi_{qlm}(\mbf R)=Y_{qlm}(\mbf{\hat R})$ \cite{jainbook}. This feature again elucidates the duality between Haldane sphere and chiral multifold fermions, where the monopole harmonics wavefunctions are also manifest in the eigenstates.

Having dualized the chiral multifold fermions onto the Haldane spheres, we now illustrate explicitly how the trace of quantum metric (\ref{eq:trqmangmomt}) acquires the quantization (\ref{eq:trqm}) for the chiral multifold fermions. The essential point is to uncover the relation between the dynamical angular momentum $\mbf\L^{sn}$ and the actual angular momentum $\mbf L^{sn}$ under rotation symmetry \cite{haldane83prl}. This can be achieved by examining the commutation relations and composing the one which satisfies the $\txt{SU}(2)$ Lie algebra. The analysis starts by calculating the commutation relation of dynamical angular momentum, yielding $[\L^{sn}_a,\L^{sn}_b]=i\ve_{abc}(\L^{sn}_c-q^{sn}\hat k_c)$. This result motivates the derivation of another commutation relation $[\L^{sn}_a,\hat k_b]=i\ve_{abc}\hat k_c$. Based on these two relations, we identify the angular momentum as $\mbf L^{sn}=\mbf \L^{sn}+q^{sn}\mbf{\hat k}$, whose commutation relation manifests the $\txt{SU}(2)$ Lie algebra $[L^{sn}_a,L^{sn}_b]=i\ve_{abc}L^{sn}_c$. A correspondence between the angular momentum and the good quantum number in the model (\ref{eq:ham0}) is then established $|\mbf L^{sn}|^2=s(s+1)$. To calculate the trace of quantum metric (\ref{eq:trqmangmomt}) in terms of the good quantum numbers $s$ and $q$, we utilize the expression $|\mbf\L^{sn}|^2=|\mbf L^{sn}-q^{sn}\mbf{\hat k}|^2$ and note that $\mbf \L^{sn}\cdot\mbf{\hat k}=\mbf{\hat k}\cdot\mbf \L^{sn}=0$. The calculation confirms the dual Landau level quantization for the trace of quantum metric (\ref{eq:trqm}). This further justifies the validity of the geometric invariant $G^{sn}$ (\ref{eq:geominv}) and the sum rule (\ref{eq:sumrule}) along with the Chern number $C^{sn}$. Despite the initial induction based on observation, the dual Haldane sphere provides a rigorous and concise derivation that solidates the results.

The quantized trace of quantum metric (\ref{eq:trqm}) and according geometric invariant (\ref{eq:geominv}) can have interesting effects on various measurable physical quantities. To study these manifestations, we assume a general 3D multiorbital system which exhibits a multiband structure and hosts the chiral multifold fermions. The model (\ref{eq:ham0}) is realized at a band crossing point $\mbf K$ in the Brillouin zone (BZ), with the eligible region $\mca R_\txt{CMF}$ defined by a radial momentum cutoff $\L_k$. Note that the spin-orbit-coupled rotation symmetry serves as an approximate symmetry in this low-energy theory. For a certain band involved in the band crossing, the Bloch state $\ket{u^n_{\mbf k}}$ is described by the eigenstate $\ket{u^n_{\mbf k}}=\ket{u^{sn}_{\mbf k}}$ of the model (\ref{eq:ham0}) in the eligible region $\mca R_\txt{CMF}$. On the linearly dispersing bands $n\neq0$, the Fermi surfaces are spherical shells at finite doping from the band crossing. Such spherical shells locate at radius $k_F=|\mu|/v|n|$, where $\mu\neq0$ is the relative chemical potential to the band crossing. Meanwhile, the flat bands $n=0$ can occur in the integer spin models $s=1,2,\dots$. The according Fermi surfaces at $\mu=0$ are solid spheres with radius $\L_k$, which covers the whole eligible region $\mca R_\txt{CMF}$ of the low-energy theory.

An important basis for the manifestations of quantum metric lies in the spread of Wannier functions \cite{marzari97prb,matsuura10prb,marzari12rmp}. Wannier functions are the localized representations of electronic states in real space. For a band $\ket{u^n_{\mbf k}}$ in the multiband structure, the Wannier function $\ket{\mbf Rn}$ at lattice vector $\mbf R$ is constructed by a Fourier transform from the Bloch states $\ket{\mbf Rn}=(\mca V_0/\mca V)\sumv{k}\intv{r}\ket{\mbf r}\innp{\mbf r}{u^n_{\mbf k}}e^{i\mbf k\cdot(\mbf r-\mbf R)}$. Here $\mca V_0$ and $\mca V$ denote the volumes of primitive unit cell and whole system, respectively. The availability of exponentially localized Wannier functions are usually expected for a single isolated band \cite{marzari12rmp}. However, such exponential localization may be lost for a single band in a set of composite bands, known as the Wannier obstruction. To capture the localization in the Wannier functions, the `spread functional' $\O^n=\braket{\mbf0n}{\mbf r^2}{\mbf0n}-\braket{\mbf0n}{\mbf r}{\mbf0n}^2$ was defined as a quantitative measure \cite{marzari97prb}. This functional contains a gauge invariant part $\O^n_I$ as a lower bound $\O^n\geq\O^n_I$, which is constant under any gauge transformation $\ket{u^n_{\mbf k}}\rar e^{i\phi^n_{\mbf k}}\ket{u^n_{\mbf k}}$. Significantly, the contribution from band geometry is encoded in the gauge invariant part of spread functional $\O^n_I=(\mca V_0/\mca V)\sumv{k}\Tr g^n_{\mbf k}$. As nontrivial band geometry occurs from the $\mbf k$-dependent orbital composition, the single band may become insufficient for exponential localization, leading to a finite spread in the Wannier function. For the bands involved in the band crossing (\ref{eq:ham0}), the contributions from the chiral multifold fermions can be further distinguished $\O^{sn}_{\txt{CMF},I}=(\mca V_0/\mca V)\sum_{\mca R_\txt{CMF},\mbf k}\Tr g^{sn}_{\mbf k}$. Note that this sets a lower bound of the spread functional $\O^{sn}_I\geq\O^{sn}_{\txt{CMF},I}$, since the trace of quantum metric is a momentum-space Laplacian and is positive semidefinite. We thus identify a lower bound of the spread functional solely from the band geometry of chiral multifold fermions. With an integration over the eligible region $\mca R_\txt{CMF}$, we determine this lower bound from the geometric invariant $G^{sn}$ (\ref{eq:geominv})
\beeq
\O^{sn}_I\geq\fr{\mca V_0\L_k}{4\pi^2}G^{sn}.
\eneq
Notably, the bands with smaller monopole charge (such as the flat trivial bands with $q^{s0}=0$) exhibit larger lower bounds for the finite spread. This feature differs remarkably from the usual understanding of Wannier obstruction, which expects a higher degree of obstruction on a band with more nontrivial topology.

As a direct consequence of Wannier obstruction from band geometry, the chiral multifold fermions can form superconductivity even if the bands are (nearly) flat \cite{lin20prr}. When a superconducting band is flattened, the Cooper pairs become nondispersive and well localized. This may lead to the lost of interpair communication, thereby suppress the phase coherence of superconductivity. A quantitative measure of phase coherence is provided by the superfluid stiffness $D^S_{ab}$, which captures the response of a supercurrent $j^S_a$ to an electromagnetic gauge field $\mbf A_b$. The scaling $D^S_{ab}\sim v_F^2$ with respect to Fermi velocity $v_F$ confirms the loss of phase coherence $D^S_{ab}\rar0$ in the flat band limit $v_F\rar0$. Due to the absence of global phase coherence, the obstruction to superconductivity is usually expected on flat bands. Nevertheless, `anomalous phase coherence' may arise and support superconductivity on a single flat band in a set of composite bands \cite{peotta15nc}. Despite the localization of Cooper pairs, the overlaps of wavefunctions from Wannier obstruction can still mediate the phase coherence. Such effect is reflected by the anomalous superfluid stiffness \cite{peotta15nc,liang17prb,hu19prl,xie20prl,lin20prr}
\beeq
\label{eq:sfstf}
D^{S,n}_{\txt{geom},ab}(T)=\fr{1}{\mca V}\sumv{k}\fr{2|\D_{\mbf k}|^2}{E^n_{\mbf k}}\tanh\fr{E^n_{\mbf k}}{2T}g^n_{ab\mbf k},
\eneq
where $\D_{\mbf k}$ is the superconducting gap function, $E^n_{\mbf k}$ is the quasiparticle energy, and $T$ is the temperature. As a simplest illustration for the chiral multifold fermions, we calculate the anomalous superfluid stiffness of a uniform superconductivity $\D_{\mbf k}=\D(T)$ on the flat bands $n=0$ \cite{lin20prr}. With the quasiparticle energy $E^n_{\mbf k}=|\D|$, a proportionality to the gap function is established at zero temperature $T=0$
\beeq
\Tr D^{S,sn}_{\txt{geom}}(0)=\fr{\L_k}{2\pi^2}|\D(0)|G^{sn}.
\eneq
This result is also valid for the linearly dispersing bands $n\neq0$ in the flat band limit $v\rar0$. Note that the geometric invariant $G^{sn}$ (\ref{eq:geominv}) serves as an important measure of the anomalous superfluid stiffness. While previous works adopted the general relation $G^{sn}\geq|C^{sn}|$ and determined a lower bound from the Chern number \cite{peotta15nc,liang17prb,hu19prl,xie20prl}, our analysis uncovers a more precise `geometric dependence' particularly for the chiral multifold fermions. Interestingly, the bands with smaller Chern numbers manifest larger anomalous superfluid stiffness, which differs significantly from the usual expectations.

With the anomalous superfluid stiffness derived, we can further estimate the critical temperature $T_c\sim\bar D^{S,sn}_{\txt{geom}}\xi$ for the flat band superconductivity \cite{emery95n,hazra19prx}. Here $\bar D^{S,sn}_{\txt{geom}}=[\prod_aD^{S,sn}_{\txt{geom},aa}(0)]^{1/3}=(\L_k/6\pi^2)|\D(0)|G^{sn}$ from rotation symmetry $D^{S,sn}_{\txt{geom},aa}=\Tr D^{S,sn}_{\txt{geom}}/3$, and an estimation of coherence length $\xi\sim\L_k^{-1}$ is utilized \cite{hazra19prx,lin20prr}. Note that the flat band pairing leads to a dramatic enhancement $|\D(0)|\sim Vn^{sn}$, where $-V<0$ is the attraction and $n^{sn}$ is the number of states in the eligible region $\mca R_\txt{CMF}$ per unit volume \cite{lin18prb,lin20prr}. The resulting critical temperature manifests a linear scaling in the interaction strength
\beeq
T_c\sim Vn^{sn}G^{sn},
\eneq
which is much higher than the conventional exponential scaling. Such dramatic enhancement is available solely because the chiral multifold fermions host nontrivial band geometry that supports anomalous phase coherence for flat band superconductivity.

Our analysis has focused on the minimal $\mbf k\cdot\mbf S$ model (\ref{eq:ham0}) of chiral multifold fermions. In general, the low-energy theory can experience various types of perturbations, which may become relevant away from the band crossing or close to a topological phase transition of eigenstates \cite{boettcher20prl}. How stable the quantized band geometry is against these perturbations serves as an interesting question to investigate. When the perturbation respects the spin-orbit-coupled rotation symmetry, such as $\d H\sim(\mbf k\cdot\mbf S)^2$ or the irreducible representation with $(l,s,j)=(2,2,0)$ \cite{lin18prb,lin20prr}, the eigenstates are unchanged. The quantized trace of quantum metric (\ref{eq:trqm}) and the geometric invariant (\ref{eq:geominv}) are still manifest, as guaranteed by the symmetry. On the other hand, the exact quantization may be obstructed when the perturbation breaks the spin-orbit-coupled rotation symmetry. Nevertheless, the original quantization can still set the characteristic scales of the nonquantized results. The surface integration $G^{sn}$ may remain close to the originally quantized value (\ref{eq:geominv}), until the structures of chiral multifold fermions disappear away from the band crossing or the divergences occur at the topological phase transitions of eigenstates \cite{ma10prb}. The study of how the band geometry varies under various perturbations is an interesting direction for future work.

Several experimental methods were proposed and realized to probe the quantum metric \cite{neupert13prb,bleu18prb,ozawa18prb,asteria19np,klees20prl,yu19nsr,tan19prl,gianfrate20n}. One of the main methods is based on the periodic drive to the system \cite{ozawa18prb}. Consider a linear shake $\d H_a=2E\cos(\o t)r_a$ with amplitude $E$, frequency $\o$, and time $t$ on a multiband system. The initial state is prepared as a Bloch state at momentum $\mbf k$ in the lowest-energy band, which may be realized by loading an according wave packet adiabatically. Significantly, the diagonal components of quantum metric can be obtained from the integrated transmission rate $\G^\txt{int}_a=2\pi E^2g_{aa\mbf k}$ under the periodic drive. Here the integrated rate reads $\G^\txt{int}_a=\int_\o\G(\o)$, and $\G(\o)$ is the time-averaged transmission probability to all of the higher-energy bands. The trace of quantum metric can thus be determined by applying the shake along all directions $a=x,y,z$ and measuring the transmission rates. Some other methods were also proposed, such as the measurement by detecting the current noise \cite{neupert13prb}. Note that the single-band quantum metric can only be probed for the lowest-energy band in all of these methods. The target of a particular band may serve as an important goal in the future development of experimental probes.

In summary, we derive an exact quantization rule of the trace of quantum metric for the chiral multifold fermions for any spin. Such derivation is achieved by dualizing the computation onto a Haldane sphere in momentum space. A surface integration around the degenerate point leads to a quantized geometric invariant. The quantized band geometry may physically manifest itself in the finite spread of Wannier functions, as well as the according anomalous phase coherence of flat band superconductivity. The quantization remains valid under the spin-orbit-coupled rotation symmetry, and may set the characteristic scales of nonquantized results under symmetry-breaking perturbations. Experimental probes with periodic drive may be adopted to obtain the quantum metric. Potential applications to the other physical observables may serve as interesting topics for future work. Meanwhile, our paradigmatic analysis may also be generalized to the other gapless topological materials, such as the nodal-loop semimetals \cite{carter12prb,weng15prx,nandkishore16prb}, the systems with 3D Z2 monopoles \cite{zhao17prl,ahn18prl}, and the topological spintronic devices \cite{ PhysRevB.95.014519}. Our analysis indicates a new framework of how the novel properties can originate from band geometry, thereby opening a new route toward the understanding of unconventional  states of matter.


\begin{acknowledgments}
The authors especially thank Dam Thanh Son and Rahul Nandkishore for encouragement and feedback on the manuscript. YPL was sponsored by the Army Research Office under Grant No. W911NF-17-1-0482. The views and conclusions contained in this document are those of the authors and should not be interpreted as representing the official policies, either expressed or implied, of the Army Research Office or the U.S. Government. The U.S. Government is authorized to reproduce and distribute reprints for Government purposes notwithstanding any copyright notation herein. WHH was supported by a Simons Investigator Grant from the Simons Foundation.
\end{acknowledgments}



\bibliography{Reference}

\end{document}